\documentclass[%
 reprint,
 amsmath,amssymb,
 aps,
]{revtex4-1}

\usepackage{graphicx}
\usepackage{dcolumn}
\usepackage{bm}
\DeclareMathOperator{\sech}{sech}

\begin{document}

\preprint{APS/123-QED}

\title{Production of Dirac Particles in External Electromagnetic Fields}
\author{Kenan Sogut}
\email{kenansogut@gmail.com}
 \author{Hilmi Yanar}%
  \author{Ali Havare}
\affiliation{%
 Mersin University, Department of Physics, 33343, Mersin, TURKEY.}%

\date{\today}

\begin{abstract}
Pair creation of spin-$\frac{1}{2}$ particles in Minkowski spacetime
is investigated by obtaining exact solutions of the Dirac equation
in the presence of electromagnetic fields and using them for
determining the Bogoliubov coefficients. The resulting particle
creation number density depends on the strength of the electric and
magnetic fields.
\begin{description}
\item[PACS numbers]
13.40.-f, 23.20 Ra.
\end{description}
\end{abstract}

\pacs{Valid PACS appear here}
\maketitle

\section{Introduction}

After the pioneering works of Sauter \cite{1}, Heisenberg and Euler
\cite{2} on the particle creation by the strong electromagnetic
fields, Schwinger formulated the following pair creation probability
per unit volume and time by obtaining the one-loop effective action
in a  constant and homogeneous classical electric field (in natural
units, $\hbar = c = 1$) \cite{3}:
\begin{equation}
\omega=\frac{(eE)^{2}}{4\pi^{3}}\sum_{n=1}^{+\infty}\frac{1}{n^{2}}exp(-\frac{n\pi
m^{2}}{eE})
\end{equation}
where $m$ and $e$ are the mass and charge of the electron, $E$ is
the electric field, respectively. Since then, this process is called
Schwinger mechanism and has become an important problem in the
quantum field theory (QFT). Such kind of a classical electric field
is assumed to be order of $E \sim 10^{16} V/cm$ \cite{4} which is
very difficult to generate by the current technology. Strong fields
arising from the collisions between the relativistic high energy
particles and heavy-ions are called color electric fields and have
ability to create particles from the vacuum. These type of
collisions are generated at the modern colliders, i.e at CERN.
Schwinger mechanism is attributed to the hadronic particle creation
and on the base of Color Glass Condensate (CGS), this phase is
called as Glasma.

The Schwinger mechanism have been studied in the presence of various stationary and non-stationary external fields \cite{5}-\cite{9}. The studies about the Schwinger mechanism in gauge fields having both electric and
magnetic field components have revealed that electric field has a dominant influence in
creating the particles. Therefore, the pair creation mechanism is
totally attributed to the pure electric field \cite {10}. This
quantum effect of the classical electromagnetic fields is carried
out to the curved spacetime as well \cite{11}-\cite{13}.

There is a considerable point in some of the studies in the
literature that they support the magnetic field has a reduction
effect in the particle creation process. One of the aims of this
study is to investigate this phenomena for a particular choice of
the electromagnetic gauge field that has both electric and magnetic
field components.

The calculation of fermionic particle creation rate requires to
define the positive and negative frequency energy states, namely the
"in" and "out" mode vacuum solutions. For the motion of the
relativistic charged particles moving in an external field, analysis
of mode functions as positive and negative frequency solutions is
not easy since the Lagrangian of the corresponding system completely
depends on space-time coordinates. Namely, particle concept becomes
indefinite owing to interaction with the external fields. For this
reason we require a condition to define "particle" concept. In the
present study we will apply a quasiclassical method. We obtain exact
solutions of the Hamilton-Jacobi (HJ) equation and discuss their
asymptotic behavior in the infinite past and future . Then,
asymptotic behavior of the solutions of the Dirac equation in the
neighborhood of the time singularities will be identified. With the
help of this analysis and comparison of asymptotic solutions of both
HJ and Dirac equations in the infinite past and future, the particle
picture will be identified.

We define positive and negative frequency mode functions in such a
way that the positive frequency mode function approaches $e^{iS(t)}$
and the negative frequency one $e^{-iS(t)}$ in asymptotic regions
\cite{4}, where $S(t)$ is the solution of the HJ equation for the
presence of a $4$-vector electromagnetic potential given as
\begin{equation}
A_{\nu}=B_{0}\tau[1+\tanh(x/\tau)]\delta_{\nu}^{2}-E_{0}(\Gamma+
\Lambda t)\delta_{\nu}^{3}
\end{equation}
where $\tau$, $\Gamma$ and $\Lambda$ are constants. This new suggested form of the
vector potential generates parallel stationary electric \cite{5} and Sauter type magnetic fields \cite{14}
that are persuaded in the Glasma flux tube model of high energy
heavy ion collisions.
\newline Magnetic current emerging is found to
be
\begin{equation}
j_{\nu}=\frac{1}{4\pi}[\frac{2|\overrightarrow{B}|}{\tau}\tanh(x/\tau)]\delta_{\nu}^{2}
\end{equation}

The outline of the paper is as follow: In Section $2$ we solve the
relativistic HJ equation and obtain the asymptotic behavior of the
solutions. In Section $3$ we solve the Dirac equation for the
considered electromagnetic fields and obtain the asymptotic limits
of the solutions to define the vacuum "in" and "out" modes by
referring the asymptotic solutions of the HJ equation. We use the
Bogoliubov transformation technique to relate the solutions at the
boundaries and calculate the particle creation number density for
fermions in Section $4$. Finally, in Section $5$ we discuss the
results we obtained. Throughout the paper the natural units, $\hbar
̄ = c = 1$ are used.

\section{Solutions of the Hamilton-Jacobi Equation}

The relativistic HJ equation for the action S is given by \cite{11}:
\begin{equation}
\zeta^{\epsilon\theta}[\frac{\partial S}{\partial
x^\epsilon}-eA_{\epsilon}][\frac{\partial S}{\partial
x^\theta}-eA_{\theta}]+m^{2}=0
\end{equation}
where $\zeta^{\epsilon\theta}=(1,-1,-1,-1)$ is the Minkowski metric,
$m$ is the mass of the particle and $A_{\nu}$ is the $4$-vector
electromagnetic potential.

The electromagnetic potential satisfy the Lorentz gauge and the
Lorentz invariants are determined from the electromagnetic field
tensor as follows
\begin{eqnarray}
F^{\pi\varrho}F_{\pi\varrho}=2({B}^{2}-{E}^{2})=2B_{0}^{2}\sech^{4}(x/\tau)-E_{0}^{2}\Lambda^{2}
\end{eqnarray}
and
\begin{equation}
F^{\pi\varrho}F^{*}_{\pi\varrho}=4\overrightarrow{E}\cdot
\overrightarrow{B}=4\Lambda E_{0}B_{0}\sech^{2}(x/\tau)
\end{equation}
Because of the space-time dependence of the considered
electromagnetic field, the solution of the HJ equation can be
separated as follow:
\begin{equation}
S(t,\overrightarrow{x})=P(x)+Q(t)+(yk_{y}+zk_{z})
\end{equation}
where $k_{y}$ and $k_{z}$ can be viewed as the conserved momenta
that exist given the symmetries chosen for the electromagnetic gauge
(2). By using (7) in Eq. (4) we obtain
\begin{eqnarray}
\dot{Q}^{2}-\acute{P}^{2}-[k_{z}+eE_{0}(\Gamma+ \Lambda t)]^{2}
\nonumber \\
-[k_{y}-eB_{0}\tau(1+\tanh(x/\tau))]^{2}+m^{2}=0
\end{eqnarray}
where dot and acute denote derivatives with respect to $t$ and $x$,
respectively.

We obtain two first order differential equations as follows:
\begin{equation}
\dot{Q}^{2}-[eE_{0}(\Gamma+ \Lambda t)]^{2}-2k_{z}eE_{0}(\Gamma+
\Lambda t)+m^{2}-k_{z}^{2}=v^{2}
\end{equation}
and
\begin{equation}
\acute{P}^{2}+\{eB_{0}\tau[1+\tanh(x/\tau)]\}^{2}-2k_{y}eB_{0}\tau[1+\tanh(x/\tau)]=v^{2}
\end{equation}
where $v^{2}$ is the constant of separation.

Time-dependent external fields cause unstable vacuum and this
results in the pair creation by vacuum. For this reason the dynamics
involving spatial coordinates effect the solutions only by a
constant and we obtain the solution of the HJ equation for
electromagnetic gauge (2) as follow
\begin{align}
S(\varrho,\overrightarrow{x})&=S_{0}(0, \overrightarrow{x})\nonumber \\
&+\frac{eE_{0}}{\Lambda}\int_{0}^{\varrho} \sqrt{\varrho^{2}+\frac{2k_{z}}{eE_{0}}{\varrho}+(\frac{k_{z}^{2}+v^{2}-m^{2}}{e^{2}E_{0}^{2}})}\, d\varrho  \nonumber\\
&=(\frac{\varrho eE_{0}+k_{z}}{2\Lambda})\sqrt{\varrho^{2}+\frac{2k_{z}}{eE_{0}}{\varrho}+(\frac{k_{z}^{2}+v^{2}-m^{2}}{e^{2}E_{0}^{2}})} \nonumber\\
&+(\frac{v^{2}-m^{2}}{2eE_{0}\Lambda})\ln\Bigg\{2\varrho+\frac{2k_{z}}{eE_{0}}\nonumber\\
&+2\sqrt{\varrho^{2}+\frac{2k_{z}}{eE_{0}}{\varrho}+(\frac{k_{z}^{2}+v^{2}-m^{2}}{e^{2}E_{0}^{2}})}\Bigg\}+S_{0}(0,\overrightarrow{x})
\end{align}
where $\varrho=(\Gamma+\Lambda t)$. \newline The dependence of the
solution on time is derived by $\psi\rightarrow e^{iS(t)}$ and we
arrive the following expressions for the asymptotic behavior of the
relativistic wave function:
\begin{equation}
\psi_{(t\rightarrow \mp\infty)}=e^{iS(t)}\rightarrow e^{\pm
i(\frac{eE_{0}\Lambda}{2})t^{2}\pm i
(\frac{v^{2}-m^{2}}{2eE_{0}\Lambda})\ln(2\Lambda|t|)}
\end{equation}
where the upper and lower signs represent the negative and
positive-frequency states, respectively.

\section{Solutions of the Dirac Equation}
\label{sec:3} The Dirac equation in external electromagnetic fields
is given by \cite{15}
\begin{equation}
[i\gamma^{\nu}\partial_{\nu}+eA_{\nu}\gamma^{\nu}-m]\psi=0
\end{equation}
where $\gamma^{\nu}$ are Dirac matrices, $A_{\nu}$ is the $4$-vector
electromagnetic potential, $m$ is the mass of electron, $e$ is the
charge of the electron and $\psi$ is the four-component spinor.

The Dirac equation yields four coupled differential equations for
the spinor and usually it is difficult to obtain the exact
analytical solutions, in particular for mathematically complicated
external fields. This difficulty of the problem has been
accomplished by Feynmann and Gell-Mann by considering a
two-component form of the Dirac equation in the presence of
electromagnetic fields as follow \cite{16}
\begin{equation}
[(\overrightarrow{P}-e\overrightarrow{A})^{2}+m^{2}-e\overrightarrow{\sigma}\cdot(\overrightarrow{B}+i\overrightarrow{E})]\phi=(p_{0}-eA_{0})^{2}\phi
\end{equation}
where $\overrightarrow{\sigma}$ are usual Pauli matrices and
$\phi=(\begin{array}{c} \phi_1 \\ \phi_2 \end{array})$ are the
solutions of the two-component equation. The four-component spinor
can be derived from $\phi$ as follow
\begin{equation}
\psi=\left(
       \begin{array}{c}
         {[{\overrightarrow{\sigma}\cdot(\overrightarrow{P}-e\overrightarrow{A})+(p_{0}-eA_{0})+m}]}\phi \\
         {[{\overrightarrow{\sigma}\cdot(\overrightarrow{P}-e\overrightarrow{A})+(p_{0}-eA_{0})-m}]}\phi \\
       \end{array}
     \right)
\end{equation}
Thence, for the purpose of obtaining the analytic solutions we
follow up the two-component formalism and consider the
electromagnetic gauge (2). Because the given gauge field depends on
$x$ coordinate and $t$, both $k_{y}$ and $k_{z}$ are constants of
the motion and solutions can be written in the form
\begin{equation}
\phi= e^{i(yk_{y}+zk_{z})}\begin{pmatrix}
           \chi_{1}(x)T_{1}(t) \\
           \chi_{2}(x)T_{2}(t)  \\
            \end{pmatrix}
\end{equation}

Therefore, with the usage of Eqs. (2) and (16), Eq.(14) becomes

\begin{eqnarray}
\{-\frac{d^{2}}{dx^{2}}-eB_{0}(eB_{0}\tau^{2}+s)\sech^{2}(x/\tau)\nonumber\\
+2eB_{0}(eB_{0}-\tau k_{y})\tanh(x/\tau) +2eB_{0}\tau(eB_{0}\tau-k_{y})\nonumber\\
+m^{2}+k_{y}^{2}+k_{z}^{2}+\frac{d^{2}}{dt^{2}}+(eE_{0}\Lambda t)^{2}+2eE_{0}\Lambda (eE_{0}\Gamma+k_{z})t \nonumber\\
+eE_{0}\Gamma (eE_{0}\Gamma+2k_{z})
-iseE_{0}\Lambda\}\chi_{s}(x)T_{s}(t)=0\nonumber\\
\end{eqnarray}
where the spin index $s$ has the $\pm1$ eigenvalues corresponding to
the spinors $\phi_{1}$ and $\phi_{2}$, respectively. This equation
can be written in a simpler form as
\begin{equation}
[\widehat{F}(x)+\widehat{Q}(t)]\chi_{s}(x)T_{s}(t)=0
\end{equation}
with the following definitions
\begin{eqnarray}
\widehat{F}(x)=-\frac{d^{2}}{dx^{2}}-eB_{0}(eB_{0}\tau^{2}+s)\sech^{2}(x/\tau)\nonumber\\
+2eB_{0}(eB_{0}-\tau k_{y})\tanh(x/\tau) \nonumber\\
+2eB_{0}\tau(eB_{0}\tau-k_{y})+m^{2}+k_{y}^{2}+k_{z}^{2}
\end{eqnarray}
\begin{eqnarray}
\widehat{Q}(t)=\frac{d^{2}}{dt^{2}}+(eE_{0}\Lambda t)^{2}+2eE_{0}\Lambda (eE_{0}\Gamma+k_{z})t \nonumber\\
eE_{0}\Gamma (eE_{0}\Gamma +2k_{z})-iseE_{0}\Lambda\
\end{eqnarray}
Eq.(18) has a separable form, so we get the following two equations
\begin{equation}
[\widehat{F}(x)-\varpi^{2}]\chi_{s}(x)=0
\end{equation}
\begin{equation}
[\widehat{Q}(t)+\varpi^{2}]T_{s}(t)=0
\end{equation}
where $\varpi^{2}$ is the constant of separation.
\newline
By defining $x=\tau r$, Eq. (21) becomes
\begin{equation}
[\frac{d^{2}}{dr^{2}}+\Sigma \sech^{2}r-\Upsilon \tanh
r-\varepsilon]\chi_{s}(x)=0
\end{equation}
where the definitions
\begin{eqnarray}
\Sigma=eB_{0}(eB_{0}\tau^{2}+s), \Upsilon=2eB_{0}\tau^{2}(eB_{0}-\tau k_{y}) \nonumber \\
\varepsilon=2eB_{0}\tau^{3}(eB_{0}\tau-k_{y})+\tau^{2}(m^{2}+k_{y}^{2}+k_{z}^{2}-\varpi^{2})
\end{eqnarray} were made.

Following Rosen and Morse \cite{17}, we set
$\chi_{s}(r)=e^{ra}\cosh^{-b}rf_{s}(r)$ and obtain the following
equation
\begin{eqnarray}
\{f^{"}_{s}+2(a-b\tanh r)f^{'}_{s}+[(\Sigma-b(b+1))\sech^{2}r \nonumber\\
-(2ab+\Upsilon)\tanh r+(a^{2}+b^{2}-\varepsilon)] f_{s}\}=0
\end{eqnarray}
In order $\chi/f$ to be finite in the range $-\infty\leq r\leq
+\infty$, $(a^{2}+b^{2}-\varepsilon)=0$ and $(2ab+\Upsilon)=0$
conditions are necessary \cite{17}. From these conditions we derive
the following expressions for $a$ and $b$:
\begin{equation}
a=-\frac{1}{2}[(\varepsilon+\Upsilon)^{\frac{1}{2}}-(\varepsilon-\Upsilon)^{\frac{1}{2}}]
\end{equation}
and
\begin{equation}
b=\frac{1}{2}[(\varepsilon+\Upsilon)^{\frac{1}{2}}+(\varepsilon-\Upsilon)^{\frac{1}{2}}]
\end{equation}

Therefore, keeping these expressions and by introducing
$\eta=\frac{1}{2}(1+\tanh r)$ we arrive
\begin{equation}
\{\eta(1-\eta)\frac{d^{2}}{d\eta^{2}}+[a+b+1-2(b+1)\eta]\frac{d}{d\eta}+[\Sigma-b(b+1)]\}f=0
\end{equation}
which is the differential equation satisfied by the hypergeometric
functions. The hypergeometric function remaining finite at $\eta=0$
will provide this equation and solution will be given as \cite{18}
\begin{eqnarray}
f_{s}(\eta)=_{2}F_{1}[(b+1/2)-(\Sigma+1/4)^{\frac{1}{2}}; \nonumber \\
(b+1/2)+(\Sigma+1/4)^{\frac{1}{2}}; a+b+1; \eta]
\end{eqnarray}
So, we obtain
\begin{eqnarray}
\chi_{s}=e^{ra}\cosh^{-b}r _{2}F_{1}[(b+1/2)-(\Sigma+1/4)^{\frac{1}{2}}; \nonumber \\
(b+1/2)+(\Sigma+1/4)^{\frac{1}{2}}; a+b+1; \eta]
\end{eqnarray}
For this solution to be convergent at infinity the following
condition must be satisfied \cite{17}
\begin{equation}
[(b+1/2)-(\Sigma+1/4)^{\frac{1}{2}}]=-n
\end{equation}
Then
\begin{equation}
a=-\Upsilon[(4 \Sigma+1)^{\frac{1}{2}}-(2n+1)]^{-1}
\end{equation}
and
\begin{equation}
b=\frac{1}{2}[(4\Sigma+1)^{\frac{1}{2}}-(2n+1)]
\end{equation}
The constant of separation $\varpi$ can be easily derived from
$(a^{2}+b^{2}-\varepsilon)=0$.

By introducing a variable
$\xi=\sqrt{\frac{2}{eE_{0}\Lambda}}(eE_{0}\Lambda
t+eE_{0}\Gamma+k_{z})$ we obtain the following equation from Eq.(22)
\begin{equation}
\{\frac{d^{2}}{d\xi^{2}}+\frac{1}{4}\xi^{2}-\frac{iesE_{0}\Lambda+k_{z}^{2}-\varpi^{2}}{2eE_{0}\Lambda}\}T_{s}(\xi)=0
\end{equation}
Solutions of this differential equation are parabolic cylinder
functions \cite{18}
\begin{equation}
T_{s}(\xi)=\frac{e^{-\frac{\pi
\tilde{a}}{4}}}{(2eE_{0}\Lambda)^{\frac{1}{4}}}[D_{-i\tilde{a}-1/2}(e^{i\pi
/4}\xi)+D_{-i\tilde{a}-1/2}^{\ast}(e^{i\pi /4}\xi)]
\end{equation}
where
$\tilde{a}=(\frac{iesE_{0}\Lambda+k_{z}^{2}-\varpi^{2}}{2eE_{0}\Lambda})$.
\newline
Therefore, exact solutions are obtained and all components of the
Dirac spinor can be found with the insertion of Eqs. (30) and (35)
into Eq.(16).

\section{Particle Creation via Bogoliubov Transformation Method}
Due to difficulty of the direct observation of the pair creation in
a constant field \cite{10}, because the typical $\vert eE \vert$ is
smaller than $m^{2}$, the particle creation will be induced by the
time-dependent components of the wave-function (28), namely the
parabolic cylinder functions.
\newline
Two solutions of the Eq. (34) are given as:
\begin{equation}
T_{s_{1}}(\xi)=\frac{e^{-\frac{\pi
\tilde{a}}{4}}}{(2eE_{0}\Lambda)^{\frac{1}{4}}}D_{-i\tilde{a}-1/2}(e^{i\pi
/4}\xi)
\end{equation}
and
\begin{equation}
T_{s_{2}}(\xi)=\frac{e^{-\frac{\pi
\tilde{a}}{4}}}{(2eE_{0}\Lambda)^{\frac{1}{4}}}D_{-i\tilde{a}-1/2}^{\ast}(e^{i\pi
/4}\xi)
\end{equation}
These are not the only solutions and any of the remaining two-sets
can be constructed via Bogoliubov coefficients as follow:
\begin{equation}
\tilde{T}_{s_{1}}(\xi)=\alpha T_{s_{1}}(\xi)-\beta^{*}T_{s_{2}}(\xi)
\end{equation}
and
\begin{equation}
\tilde{T}_{s_{2}}(\xi)=\alpha^{*} T_{s_{2}}(\xi)+\beta
T_{s_{1}}(\xi)
\end{equation}
\newline
The Bogoliubov transformation method is a technique that associates
a canonical commutation relation algebra or a canonical
anti-commutation relation algebra into another representation,
caused by an isomorphism \cite{19}.
\newline
In the Minkowskian QFT, eigenfunctions of the field equation,
$\psi$, can be written with the help of the mode solutions as
\cite{19}-\cite{20}.
\begin{equation}
\psi=\sum_{n}(a_n\varphi_n+ a_n^\dag\varphi_n^\ast )
=\sum_{k}(b_k\Theta_k+ b_k^\dag\Theta_k^\ast )
\end{equation}
where we have the relations $(\varphi_{i},\varphi_{j})=\delta_{ij}$,
$(\varphi_{i}^{\ast},\varphi_{j}^{\ast})=\delta_{ij}$,
$(\varphi_{i},\varphi_{j}^{\ast})=0$ and
$(\Theta_{i},\Theta_{j})=\delta_{ij}$,
$(\Theta_{i}^{\ast},\Theta_{j}^{\ast})=\delta_{ij}$,
$(\Theta_{i},\Theta_{j}^{\ast})=0$ for $\varphi$ and $\Theta$ are
mode solutions. The $\varphi$ and $\Theta$ can be expanded in terms
of each other.

The creation and annihilation operators $a_n^\dag, b_k^\dag $ and
$a_n, b_k $ are in correlation by the following expressions
\begin{equation}
a_n=\sum_{k}(\alpha_{kn} b_k+ \beta_{kn}^\ast b_k^\dag )
\end{equation}
\begin{equation}
b_k=\sum_{n}(\alpha_{kn}^\ast a_n - \beta_{kn}^\ast a_n^\dag )
\end{equation}
$\alpha_{kn}$ and $\beta_{kn}$ are Bogoliubov coefficients
determined by $\alpha_{ij}=(\Theta_{i},\varphi_{j})$,
$\beta_{ij}=-(\Theta_{i},\varphi_{j}^{\ast})$. They are related as
\begin{equation}
\sum_{i}(\alpha_{ni} \alpha_{ki}^\ast - \beta_{ni}
\beta_{ki}^\ast)=\delta_{nk}
\end{equation}
\begin{equation}
\sum_{i}(\alpha_{ni} \beta_{ki} - \beta_{ni} \alpha_{ki})=0.
\end{equation}
Let $|0_{a}\rangle$ and $|0_{b}\rangle$, are two states of vacuum in
the Fock space and are related to each particle notion in (30). They
are represented for all $n$ and $k$ as
\begin{equation}
|0_{a}\rangle :  a_n |0_{a}\rangle=0
\end{equation}
\begin{equation}
|0_{b}\rangle :  b_k |0_{b}\rangle=0
\end{equation}
If $|0_{b}\rangle$ is introduced as the usual vacuum, then
$|0_{a}\rangle$ is regarded as a many-particle state. Therefore, the
number of $\Theta_{n}$-mode particles in the state of
$|0_{a}\rangle$ is
\begin{equation}
\langle 0_{a}|b_k^\dag b_k| 0_{a}\rangle=\sum_{n}|\beta_{kn}|^2
\end{equation}
If the $\varphi_{n}(x)$ are defined as positive frequency modes and
the $\Theta_{n}(x)$ modes are linear unification of them, then
$\beta_{jk}=0$. Then, $b_k |0_{b}\rangle=0$ and $a_k
|0_{a}\rangle=0$. Hence, $\varphi_{j}$ and $\Theta_{k}$ modes have a
common vacuum state. If $\beta_{jk}\neq0$, then $\Theta_{k}$ contain
a combination of positive-$\varphi_{k}$ and
negative-$\varphi_{k}^{\ast}$ frequency modes.

Therefore, we can define the positive- and negative-frequency
solutions in order to find the Bogoliubov coefficients. Asymptotic
expansion of the parabolic cylinder functions is gives by \cite{21}
\begin{equation}
D_{\nu}(z)_{|z|\rightarrow +\infty}\approx z^{\nu}e^{-z^{2}/4} ,
|arg z|<\frac{3\pi}{4}
\end{equation}
Taking into account this relation for Eqs. (36),(37) in the limit
$t\rightarrow +\infty$ (namely, $\xi \rightarrow +\infty$) and
comparing the asymptotic expansion of them with Eq. (12), we see
that the positive and negative-frequency mode solutions will be as
follows respectively,
\begin{equation}
T_{s_{1}}(\xi)\approx (\sqrt{2eE_{0}\Lambda}|t|)^{-1/2}
e^{(-ieE_{0}\Lambda t^{2}/2
-i\tilde{a}\ln(\sqrt{2eE_{0}\Lambda}|t|))}
\end{equation}
and
\begin{equation}
T_{s_{2}}(\xi)\approx (\sqrt{2eE_{0}\Lambda}|t|)^{-1/2}
e^{(ieE_{0}\Lambda t^{2}/2
+i\tilde{a}\ln(\sqrt{2eE_{0}\Lambda}|t|))}
\end{equation}
We conclude that the solutions behave as $T_{\pm} \approx e^{\pm
iS(t)}$.
\newline
For $t\rightarrow -\infty$($\xi \rightarrow -\infty$), the solutions
are in the form
\begin{equation}
T_{s_{1}}(\xi)=\frac{e^{-\frac{\pi
\tilde{a}}{4}}}{(2eE_{0}\Lambda)^{\frac{1}{4}}}D^{*}_{-i\tilde{a}-1/2}(-e^{i\pi
/4}\xi)
\end{equation}
and
\begin{equation}
T_{s_{2}}(\xi)=\frac{e^{-\frac{\pi
\tilde{a}}{4}}}{(2eE_{0}\Lambda)^{\frac{1}{4}}}D_{-i\tilde{a}-1/2}(-e^{i\pi
/4}\xi)
\end{equation}
so that their asymptotic behavior should be $T_{\pm} \approx e^{\pm
iS(t)}$. It is clear that the solutions are different in the
asymptotic regions and this is the nature of the particle creation.
Therefore, the solutions for $t\rightarrow +\infty$ belong to vacuum
"out" mode whereas are vacuum "in" mode for $t\rightarrow -\infty$.
\newline The positive and negative frequency vacuum "out" and "in" modes
can be related to each other with the Bogoliubov coefficients. By
using Eq. (39), we can write
\begin{eqnarray}
D_{-i\tilde{a}-1/2}(-e^{i\pi
/4}\xi)=\alpha^{\ast}D^{\ast}_{-i\tilde{a}-1/2}(e^{i\pi /4}\xi)
\nonumber \\
+\beta D_{-i\tilde{a}-1/2}(e^{i\pi /4}\xi)
\end{eqnarray}
Expanding the left side of this expression according to the below
formula \cite{21}
\begin{equation}
D_{\nu}(z)=
[e^{-i\pi\nu}D_{\nu}(-z)+\frac{\sqrt{2\pi}}{\Gamma(-\nu)}e^{-i\pi
(\nu+1)/2}D_{-\nu-1}(iz)]
\end{equation}
and using the result $[D_{\nu}(z)]^{\ast}=D_{-\nu-1}(-iz)$ that can
be derived easily by taking the advantage of the relation between
the parabolic cylinder function and Whittaker function given as
\cite{21}
\begin{equation}
D_{\nu}(z)=2^{(\nu+\frac{1}{2})/2}z^{-1/2}W_{\frac{1}{2}(\nu+\frac{1}{2}),-\frac{1}{4}}(z^{2}/2)
\end{equation}
we obtain the Bogoliubov coefficients $\alpha$ and $\beta$ as
follows:
\begin{equation}
\alpha=\frac{\sqrt{\frac{2\pi}{\breve{a}}}ie^{-\pi
\breve{a}/2}}{\Gamma(-i \breve{a})}
\end{equation}
and
\begin{equation}
\beta=e^{-\pi \breve{a}}
\end{equation}
where $\breve{a}=(\frac{k_{z}^{2}-\varpi^{2}}{2eE_{0}\Lambda})$ and
$|\alpha|^{2}+|\beta|^{2}=1$ condition is satisfied.
\newline Then, we find the below expression for the Bogoliubov
coefficients
\begin{equation}
\frac{\vert\alpha\vert^{2}}{\vert \beta
\vert^{2}}=\frac{2\pi}{\breve{a}}e^{\pi
\breve{a}}\frac{1}{|\Gamma(-i \breve{a})|^{2}}
\end{equation}
\newline
By considering the following formula for Gamma functions \cite{17}
\begin{equation}
{\vert \Gamma (iq ) \vert^{2}}=\frac{\pi}{q \sinh (\pi q)}
\end{equation}
the number density of the created particles can be computed as
follow
\begin{equation}
N\simeq \vert\beta\vert^{2}=[\frac{\vert\alpha\vert^{2}}{\vert \beta
\vert^{2}}+1]^{-1}=e^{-2\pi\breve{a}}
\end{equation}
where the parameter $\breve{a}$ in terms of the physical constants of four-vector potential (2) has been given as below.
\begin{equation}
\begin{aligned}
\breve{a}={} & \frac{1}{2eE_{0}\Lambda}[\frac{4e\tau^{2}B_{0}^{2}(eB_{0}-\tau k_{y})^{2}}{\left(-1-2n+\sqrt{1+4eB_{0}(s+eB_{0}\tau^{2})}\right)^{2}}\\ 
             &-\frac{1}{4\tau^{2}}\left(-1-2n+\sqrt{1+4eB_{0}(s+eB_{0}\tau^{2})}\right)^{2}\\ 
             &-(m^{2}+k_{y}^{2})-2eB_{0}\tau(eB_{0}\tau-k_{y})]
\end{aligned}
\end{equation}

\section{Conclusion}
In this study, we used the two-component formalism for the Dirac
equation that is proposed by Feynmann and Gell-Mann. This approach
to the problem removes the complexity of obtaining the exact
solutions. One of the advantages of working with this form of the
Dirac equation is that these solutions are valid for the
Klein-Gordon particles in the case of $s = 0$. Thus the results can
be used both for scalar and fermionic particles.

Mechanism of particle production by strong electric fields is
significant in order to figure out the  early stages of the
heavy-ion collisions, for example their effect on the thermalization
of quarks and gluons. For the analysis of our problem we take
account a strong constant electric field and a space-dependent
hyperbolic magnetic field. Exact solutions of the Dirac equation
were identified in terms of the parabolic cylinder and
hypergeometric functions.

\begin{figure}[h]
\begin{center}
   \includegraphics[scale=0.75]{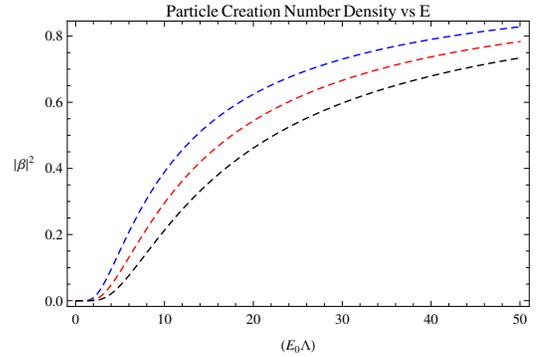}
     \caption{Particle creation number density versus electric field strength is depicted. $m=1; n=1; \tau=1; k_y=1; k_z=1; \Lambda=1$ and $B_{0}:0 (blue)$, $B_{0}:0.2 (red)$, $B_{0}:0.4 (black)$}
\end{center}
   \end{figure}
Existence of the strong electric fields cause to unstable vacuum
that is asymptotically static at future. The "in" and "out" vacuum
states were determined with the help of the asymptotic solutions of
relativistic HJ equation. They were related by the Bogoliubov
coefficients that are used to calculate the particle creation number
density in Eq.(60). This expression depends on the parameters of
electric and magnetic fields and is not in Fermi-Dirac thermal form.
As it is seen by analyzing the formula and also from the Figure $1$,
selected form of the magnetic field has a reduction effect on the
creation of fermionic particles. This situation is compatible with
previous obtained results. Also it can be seen from Figure $1$,
particle creation rate increases due to electric field strength, $(E_{0}\Lambda)$.

\begin{acknowledgments}
This study is supported by the Research Fund of Mersin University in
TURKEY with project number: 2016-1-AP4-1425.
\end{acknowledgments}

\end{document}